\documentstyle[12pt,epsfig]{article}
\begin{document}
\def\scri{\unitlength=1.00mm
\thinlines
\begin{picture}(3.5,2.5)(3,3.8)
\put(4.9,5.12){\makebox(0,0)[cc]{$\cal J$}}
\bezier{20}(6.27,5.87)(3.93,4.60)(4.23,5.73)
\end{picture}}

\begin{center}

{\Large ANTI-DE SITTER QUOTIENTS:}

\vspace{10mm}

{\Large  WHEN ARE THEY BLACK HOLES?}

\vspace{15mm}

{\large Stefan \AA minneborg}\footnote{Email address: 
stefan.aminneborg@utbildning.stockholm.se}

\vspace{5mm}

{\sl Norra Reals Gymnasium\\
S-113 55 Stockholm, Sweden}

\vspace{1cm}

{\large Ingemar Bengtsson}\footnote{Email address: ingemar@physto.se. 
Supported by VR.}

\vspace{5mm}

{\sl Stockholm University, AlbaNova\\
Fysikum\\
S-106 91 Stockholm, Sweden}

\vspace{15mm}

{\bf Abstract}

\end{center}

\noindent We point out that the BTZ black holes, and their relatives, can be 
defined in a cleaner way than they originally were. The covering space can be 
taken to be anti-de Sitter space, period, while \scri \ splits up into components 
due to Misner singularities. Our definition permits us to choose 
between two conflicting claims concerning BTZ black holes in 3 + 1 dimensions. 

\newpage

{\bf 1. Introduction}

\vspace{5mm} 

\noindent Because of the special properties of the conformal boundary of 
anti-de Sitter space, it is possible to form black hole spacetimes by 
taking the quotient of anti-de Sitter space with some discrete isometry groups. 
This was first realized in the context of 2+1 dimensional gravity, and a 
black hole family with properties analogous to those of the Kerr family 
was found \cite{BTZ, BHTZ}. The construction was later generalized to more 
involved topologies \cite{brill1, Dieter, Kirill}, to higher dimensions 
\cite{4d, BLP, Eric}, and to the Kaluza-Klein case \cite{figofarr}. We call all 
black holes of this kind ``BTZ black holes''. They have attracted 
a considerable amount of attention; apparently the original paper has been 
referred to in more than 1200 publications. We will be concerned with two 
conflicting claims concerning 3+1 dimensional BTZ black 
holes: Figueroa-O'Farrill et al. \cite{figofarr} argue that a certain class 
of spacetimes contain black holes, while Holst and Peld\'an argue that they 
do not \cite{holstpeldan}. Clearly either one of these claims is wrong, 
or---the 1200 publications notwithstanding---the BTZ black holes have not 
been properly defined. 
We will argue that the second alternative holds, and will sharpen the 
definition to ensure that only one of the two claims survive.   

The appropriate definition of a black hole is as a region of spacetime which 
cannot be seen from far away; technically from its conformal boundary \scri . 
True, there may be situations where this definition fails, but for 
the simple spacetimes considered here we trust it completely. In section 
2 of this paper we discuss the conformal boundary of anti-de Sitter space, 
and stress the fact that this boundary is singular at past and future 
infinity \cite{Valentina, Tod}. Then we define the BTZ black holes. We choose 
their covering spaces to be anti-de Sitter space, with nothing removed. 
Whether they are black holes or not is decided by our treatment of their 
boundaries. In section 3 we discuss the original BTZ-Kerr black holes; 
according to our definition they are---when they spin---examples of 
singularity free and causally 
complete spacetimes, with all closed trapped surfaces and chronology violations 
confined to a black hole. (This is not quite in line with how BTZ black holes 
are usually described; in this sense we say something slightly new here.) 
In section 4 we show how the present viewpoint makes it easy to choose 
between the two conflicting claims mentioned above. 
Section 5 gives a brief summary of our argument.    

One more thing: the ``topological'' black holes \cite{4d} acquire much 
of their interest as members of a one 
parameter family of asymptotically anti-de Sitter black holes \cite{BLP}.  
The ``toroidal'' black hole to be discussed below also belongs to a 
one parameter family of spacetimes, viz. the Ehlers-Kundt B1 metrics 
\cite{Ehlers} (or ``bubbles of nothing'' \cite{Aharony}), but these 
spacetimes do not describe black holes except in one exceptional case. 
Both families of spacetimes can be obtained (locally) through analytic 
continuation from the Schwarzschild-anti-de Sitter spacetimes. Actually 
there is a third way to perform the analytic continuation, so that the 
toroidal black hole does appear as a member of a one parameter family 
of black holes. This will be described in a separate publication 
\cite{Jan}.  

\vspace{1cm}

{\bf 2. BTZ black holes defined}

\vspace{5mm}

\noindent For us, anti-de Sitter space (or adS) is a simply connected spacetime with 
constant non-zero curvature. It is the covering space of the quadric surface 

\begin{equation} X^2 + Y^2 + Z^2 - U^2 - V^2 = - 1 \end{equation}

\noindent in a flat space with the metric 

\begin{equation} ds^2 = dX^2 + dY^2 + dZ^2 - dU^2 - dV^2 \ . \end{equation} 

\noindent If the dimension differs from 3+1, adjust the number of ``spacelike'' 
coordinates. To understand the intrinsic geometry of this quadric, introduce 
an intrinsic time coordinate $t$ through 

\begin{equation} U = T\cos{t} \hspace{8mm} V = T\sin{t} \ , \hspace{8mm} 
- \infty < t < \infty \ , \hspace{5mm} T \geq 1 \end{equation}

\noindent At constant $t$ we then have one sheet of a spacelike hyperboloid 
embedded in a Minkowski space. It is well known that the intrinsic geometry 
of such a surface has constant negative curvature. 
We choose the Poincar\'e model to describe this space, that is to say 
hyperbolic space is represented as the interior of the unit ball, and its 
geodesics are arcs of circles meeting the boundary at right angles. 

In 3+1 dimensions the intrinsic metric is 
 
\begin{equation} ds^2 = - \left( \frac{1 + \rho^2}{1-\rho^2}\right)^2 dt^2 + 
\frac{4}{(1-\rho^2 )^2}\left( d\rho^2 + \rho^2(d\theta^2 + \sin^2{\theta}d\phi^2)
\right) \ , \label{salami} \end{equation}

\noindent where 

\begin{equation} 0 \leq \rho < 1 \ , \hspace{8mm} 0 < \theta < \pi \ , 
\hspace{8mm} 0 \leq \phi < 2\pi \ . \end{equation}

\noindent The 2+1 dimensional version is obtained by setting $Z = 0 \Leftrightarrow 
\theta = \pi/2$; this is an infinite salami whose slices are Poincar\'e disks.

As we have learned from Penrose, we can attach a timelike conformal boundary to 
anti-de Sitter space. We do this by a conformal rescaling of the metric; we 
will actually work with  

\begin{equation} d\hat{s}^2 = - dt^2 + \frac{4}{(1+\rho^2)^2}
\left( d\rho^2 + \rho^2\frac{4r^2}{(1+r^2)^2}(dr^2 + r^2d\phi^2)\right) \ . 
\label{ghatt} \end{equation}
   
\noindent The conformal boundary is called scri, is denoted by a script 
\scri , and is attached to the spacetime equipped with the 
unphysical metric at $\rho = 1$. Conformally the boundary is an Einstein 
universe, with dimension one less than we started out with, and with topology 
equal to a sphere cross the real line. 

In eq. (\ref{ghatt}) we somewhat incidentally introduced stereographic 
coordinates on the 2-sphere ($0 \leq r \leq \infty$). We use them to 
visualize \scri . The picture of 2+1 dimensional anti-de Sitter space 
itself is an infinite solid cylinder, with a definite choice 
of conformal structure. The picture of the 2+1 dimensional boundary of a 
3+1 dimensional anti-de Sitter space consists of two identical copies of this 
very cylinder, with their 1+1 dimensional boundaries identified, and with 
the same choice of conformal structure (so light propagates in the same way 
in the two pictures). This is depicted, with some additional detail, in fig. 
\ref{fig:ett} below. Two disks at constant time serve 
to represent a sphere; recall that a sphere can be represented by two disks, 
each of which is conformal to a hyperbolic plane. Such pictures are useful because 
they can to a large extent replace analytic calculations, once one has learned 
to recognize light cones, Killing vector flows, and the like. This information 
can be found by browsing the literature \cite{Dieter, 4d, holstpeldan, spin}. 
  
The Einstein universe is non-compact, so anti-de Sitter space is not yet 
conformally compactified; spacelike and null 
geodesics end at \scri \ , but where do the timelike geodesics go? In Minkowski 
space this question has a clear answer, but anti-de Sitter space is different. 
To see what the problem is let us introduce a time coordinate whose range is 
bounded from below, through

\begin{equation} t = \ln{\tau} \ , \hspace{8mm} 0 < \tau < \infty \ . \end{equation}

\noindent We would like to attach a regular point $i^-$ at $\tau = 0$, smoothly 
connected to \scri . 
If we also introduce a conformal factor $\Omega(\tau)$ in front of the 
metric (\ref{ghatt}), and restrict it to \scri , it becomes 

\begin{equation} d\hat{\hat{s}}^2_{|_{\rho = 1}} = 
\Omega^2(\tau) \left( - \frac{d\tau^2}{\tau^2} 
+ d\theta^2 + \sin^2{\theta}d\phi^2 \right) = - d\tau^2 + \tau^2(d\theta^2 + 
\sin^2{\theta}d\phi^2) \ , \end{equation}

\noindent where we used $\Omega = \tau$ in the second step. But this metric has 
a curvature singularity at $\tau = 0$, and so does the corresponding spacetime 
metric. The attempt to attach a point at $\tau = 0$ has therefore failed. 

It is instructive to look at the problem in the 2+1 dimensional case \cite{Valentina}. 
In this case 
our manipulations have turned \scri \ into a flat Lorentzian cylinder, with metric 

\begin{equation} d\hat{\hat{s}}^2_{|_{\rho = 1}} = - d\tau^2 + \tau^2d\phi^2 
\ , \hspace{8mm} 0 \leq \phi < 2\pi \ . \end{equation}

\noindent The covering space is a quadrant in a flat Minkowski space, but we cannot 
unroll the cylinder without destroying the topology. Hence there is of necessity 
a Misner singularity at the origin. Considered as a two dimensional spacetime 
\scri \ does not admit a maximal analytic extension to a complete Hausdorff 
manifold \cite{Misner, bible}.  

Was this conclusion an artefact of the particular way in which we tried to do 
the conformal compactification? As far as we know this point has not received much 
attention in the literature, but H. Friedrich has kindly sketched 
an argument for us, which shows that it is in fact impossible to add regular 
points at $i^-$ and $i^+$ in a smooth manner in the anti-de Sitter case. The 
argument employs congruences of timelike conformal geodesics \cite{SF}; the 
idea is that their preferred parameters must 
go through an infinite number of poles, so that timelike geodesics in anti-de 
Sitter space are infinitely long also in a conformal sense \cite{Helmut, Tod}.  
This is related to the fact that if we insist on using coordinates for which 
the metric is manifestly conformally flat, we will need an infinite number of 
patches to cover a timelike geodesic. The argument fails in 1+1 dimensions because 
the conformal geodesics are not defined there.

A proper Penrose diagram of \scri , or equivalently of the Einstein universe, 
can now be drawn. The precise form of its timelike boundaries varies depending 
on the choice of a conformal factor; our choice differs a little from the 
standard picture \cite{Tipler, Jose}. The key unnegotiable points 
are that a light ray must be able to wind an infinite number of times around it, 
and that the diagram ends with singular points both to the future and to the past. 

\begin{figure}
        \centerline{ \hbox{
                \epsfig{figure=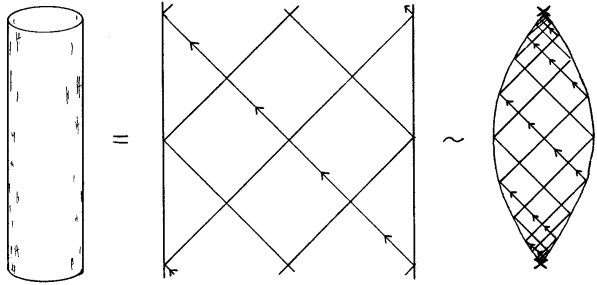,width=8cm}}}
        \caption{\small A 1+1 dimensional Einstein universe (left), unrolled to a 
strip (middle), and its conformal diagram including the singular points at 
past and future infinity (right).}
        \label{fig:4}
\end{figure}

Finally we can state a definition of a BTZ black hole spacetime. 
We begin with three requirements that we think the conformal 
boundary has to meet, in order for a $d$-dimensional spacetime to qualify 
as asymptotically anti-de Sitter: 

\

\noindent 1. \scri \ has topology ${\bf R}\times \Sigma$, where $\Sigma$ is a compact 
($d-2$)-manifold.

\noindent 2. \scri \ is locally conformal to the Einstein universe.

\noindent 3. \scri \ does not admit a conformal compactification at past and future 

temporal infinity.

\

\noindent There are some further technical requirements, for which we refer to 
Ashtekar and Magnon 
\cite{Ashtekar}. Anti-de Sitter space itself admits such a \scri , and moreover in 
this case $\Sigma$ is a sphere. We keep things more general because we will be 
forced to later on. Our definition of a BTZ black hole spacetime then becomes:    

\

\noindent 
\underline{Definition}: A BTZ black hole is a spacetime of the form adS/$\Gamma$, 
where $\Gamma$ is a discrete subgroup of the anti-de Sitter group, such that its 
conformal boundary \scri \ contains a component with the above three 
properties, and such that there is a black hole (in the ordinary sense) with 
respect to that component. 

\

\noindent Note carefully that we do not assume that the conformal boundary 
of the quotient is the quotient of the conformal boundary of anti-de Sitter 
space. Indeed the isometries we use to take the quotient will have fixed 
points on the latter, and this will 
dictate precisely how the former splits up into several components. 

In the original paper it was argued that a slightly different 
definition is to be preferred; the covering space was taken to be an open set 
of anti-de Sitter space chosen in such a way that no closed timelike curves 
appear when the quotient is taken \cite{BHTZ}. But we argue in 
section 3 that this approach weakens the analogy to the Kerr black 
hole, and in section 4 that it has led to confusion.  

\vspace{1cm}  

{\bf 3. The (2+1) Kerr family}

\vspace{5mm}

\noindent In the 2+1 dimensional case the definition of a BTZ black hole can be 
made even cleaner. Assumptions 1-3 above can be replaced by:  

\

\noindent 1$^\prime$. \scri \ contains a component conformal to the anti-de 
Sitter \scri . 

\

\noindent In this paper we will quotient with isometry groups $\Gamma$ obtained 
by exponentiating a single Killing vector field $\xi$. Here an appropriate choice 
is \cite{BHTZ} 

\begin{equation} \xi = r_+J_{XU} + r_-J_{YV} \equiv r_+(X\partial_U + 
U\partial_X) + r_-(Y\partial_V + V\partial_Y)\ , \label{xiemb} \end{equation}

\noindent where $r_+$ and $r_-$ are parameters related to mass and spin; we assume 
that $r_+ > r_-$. Inside anti-de Sitter space itself this Killing vector field will 
have fixed points if and only if $r_- = 0$, which is the spinless case. 
In the spinning case there are no fixed points, and the quotient space 
adS$/\Gamma$ is everywhere well behaved as a manifold. 

The expression (\ref{xiemb}) is not valid on \scri , but can be reexpressed in 
terms of the intrinsic coordinates used in eq. (\ref{salami}). We introduce 
the light cone coordinates

\begin{equation} u = t - \phi \hspace{8mm} v = t + \phi \ . \end{equation}

\noindent On \scri \ we obtain   

\begin{equation} \xi_{|\rho = 1} = - (r_+ + r_-)\sin{u}\partial_u - 
(r_+ - r_-)\sin{v}\partial_v \ . \end{equation}  

\noindent As explained elsewhere \cite{spin}, it is then easy to see what 
the flow lines look like, and it is also easy to see what happens when 
identifications are made along them in order to produce 
the conformal boundary of the BTZ-Kerr 
black hole. Comparing to the Penrose diagram of \scri \ it is immediately 
clear that the identifications, when applied to one of the diamonds where 
$\xi$ is spacelike, will produce a conformal copy of the anti-de Sitter 
\scri , complete with Misner singularities at the ends. Each diamond gives 
rise to separate component of \scri . In this sense the 
BTZ-Kerr spacetime is asymptotically anti-de Sitter, and a black hole 
spacetime according to our definition. 

\begin{figure}
        \centerline{ \hbox{
               \epsfig{figure=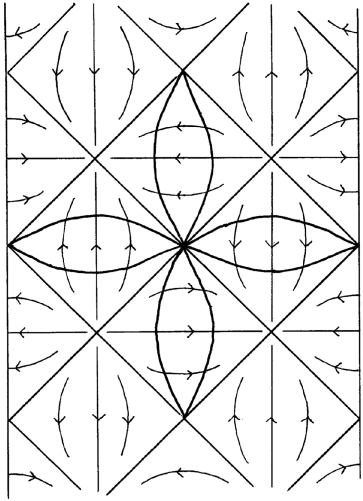,width=5cm}}}
        \caption{\small The flow of $\xi$ on a piece of the conformal 
boundary, and fundamental regions for four 
components of the conformal boundary of the quotient space. We show the spinless case; 
the spinning case is depicted elsewhere \cite{spin}.}
        \label{fig:6}
\end{figure}
 
The Misner singularities on \scri \ deserve some further comments though. 
The discrete group generated by 
$e^{\xi}$ does act properly discontinuosly on a region larger than 
the diamond shaped one that we have chosen as our covering space, say 
on two neighbouring diamonds, one where $\xi$ is spacelike and one where it is 
timelike (including their null boundary). Is it then reasonable to limit 
ourselves to one diamond only? We claim it is. First of all it is impossible 
to define a covering space leading to a geodesically complete quotient 
manifold. Secondly, when we add a neighbouring diamond to the one we started 
out from we are in the situation of Buridan's ass; we get no help to choose 
which one of two possible diamonds we should add. Including both would lead 
to a mildly non-Hausdorff geodesically incomplete manifold \cite{bible}. 
Thirdly, and in our opinion decisively, we did aim to get a conformal 
copy of the anti-de Sitter boundary as our \scri , and this is precisely 
what we do obtain when we limit ourselves to one diamond where the flow 
of $\xi$ is spacelike. Observe also that there is very little arbitrariness 
involved, since our definition 
requires \scri \ to end with a singularity to the future. For quotient black 
holes such a singularity can arise only because the isometries have fixed 
points on \scri. 

We turn to the physical spacetime. It is what it is, 
and it is not reasonable to restrict its covering space by hand. This means 
that there will be closed timelike curves in the interior of the black hole, 
coming from the regions where the flow of $\xi$ is timelike. But then it 
is precisely because of this feature that the BTZ-Kerr black hole is a very 
good analogy to the 3+1 
dimensional Kerr black hole---this also has closed timelike curves in its 
interior, as well as alternating asymptotic regions with and without them. 
Moreover the Kerr black hole just barely fails to be geodesically complete: 
causal geodesics are affected by the singularity only when confined to the 
equator \cite{boyer, carter}. This was stressed by Newman, who gave a 
discussion of geodesically complete black hole spacetimes \cite{claudel}. 
The Schwarzschild solution works in a different way, and so does the spinless 
BTZ black hole. In this case one may contemplate restricting the covering 
space in order to avoid Misner singularities; the decision affects only 
the interior of the black hole and need not concern us here.

Ba\~nados et al. \cite{BHTZ} defined their spacetimes in a different way. 
They suggested that the covering space be restricted to the open set in anti-de 
Sitter space where the flow of $\xi$ is spacelike. This evidently leads to 
an incomplete quotient space. They drew Penrose diagrams for both cases, 
but ended up advocating that the incomplete version 
is preferred. The causal structure of the BTZ-Kerr black hole in their version 
is closely analogous to that of the Reissner-Nordstr\"om black hole; it has 
a timelike singularity (or ``edge'') surrounded by outer and inner horizons, 
and is free of closed timelike curves. Now it might be argued that the 
behaviour inside the 
inner (or even outer) horizon is irrelevant, because this part of the 
solution will be changed by generic perturbations. This may be so, but 
any instability is likely to occur already at the inner horizon so we 
do not feel that this is an excuse for introducing a singularity by 
hand somewhere else. 

\vspace{1cm}

{\bf 4. Black holes and time machines in (3+1) dimensions}

\vspace{5mm} 

\noindent In 3+1 dimensions Holst and Peld\'an classified all possible spacetimes 
that arise by performing identifications using one-parameter subgroups of $SO(3,2)$ 
\cite{holstpeldan}. A black hole is obtained when the subgroup is generated by 
the Killing vector 

\begin{equation} \xi = J_{ZU} \ . \end{equation}

\noindent To see this, begin by considering the flow of $\xi$ on \scri \ , as 
sketched in fig. \ref{fig:ett}. The flow is spacelike in the region surrounding 
the pair of light cones in the cylinders depicted in fig. \ref{fig:ett}. There 
are two 
circles of fixed points at $t = \pm \pi/2$. When the identifications 
are carried through the circles of fixed points become Misner singularities in the 
quotient space, and in accordance with our philosophy we regard the region outside 
the light cones as the covering space for a component of the quotient \scri . This 
component is indeed singular (in the Misner sense) at future and past infinity; and 
there will be regions of spacetime that cannot be seen from it. Hence we do have a 
BTZ black hole \cite{4d, holstpeldan}.  

\begin{figure}
        \centerline{ \hbox{
                \epsfig{figure=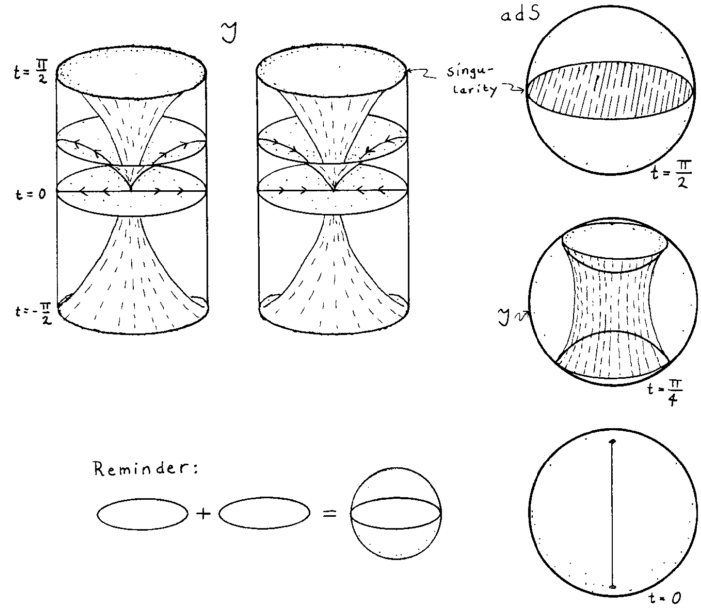,width=10cm}}}
        \caption{\small The two solid cylinders represent the conformal 
boundary of 3+1 dimensional adS. The region where $J_{ZU}$ is spacelike 
surrounds the two light cones, and 
the flow of $J_{ZU}$ defines the spacelike surfaces shown; they are punctured 
spheres, and turn into tori when the identification is carried through. The 
three balls represent three 
equal $t$-slices through anti-de Sitter space itself. They are bounded by 
spheres, appearing as constant $t$-surfaces at infinity. We remind the reader 
that the two descriptions go together because two disks can be made to represent 
a sphere. In the Poincar\'e balls we see how the event horizon is born as a 
circle at $t=0$, and then becomes a torus---that is, once the identifications 
have been carried through.}
        \label{fig:ett}
\end{figure}

Holst and Peld\'an classified all spacetimes of the form adS/$\Gamma$, 
where the discrete group $\Gamma$ is generated by a single element of the isometry group. 
One of the cases they called type Id. The generator of $\Gamma$ is then obtained 
by exponentiating the Killing vector field 

\begin{equation} \tilde{\xi} = J_{ZU} + aJ_{XY} \ , \hspace{5mm} a \neq 0 \ . \end{equation}
 
\noindent When $a = 0$ the type is Ib. They argued that, in 3+1 dimensions, type Ib is 
the only one leading to a black hole. This was disputed by Figueroa-O'Farrill et al. 
\cite{figofarr}, who argued that all spacetimes of type Id contain black holes. 
The confusion arises because of ambiguities in the choice of covering space. Holst 
and Peld\'an followed Ba\~nados et al. \cite{BHTZ} in choosing their covering space 
to be the open subset of 
anti-de Sitter space where $\tilde{\xi}$ is spacelike. With the 
coordinates and conformal factor that we use on \scri , we find 

\begin{equation} || \tilde{\xi} ||^2_{\scri} = \frac{1}{(1+r^2)^2}\left( 4r^2(\cos^2{t} 
+ a^2) - (1-r^2)^2\sin^2{t}\right) \ . \end{equation}

\noindent With increasing $a$ the region where $\tilde{\xi} $ is spacelike grows. 
As soon as $a \neq 0$ it is bounded by a timelike surface, and---crucially---it 
becomes connected. Hence the flow is spacelike in an open region including the 
boundaries of the two solid cylinders in Fig. \ref{fig:ett} for all times $t$, all 
points in the anti-de Sitter interior can be seen 
from it, and there is no black hole. 

\begin{figure}
        \centerline{ \hbox{
                \epsfig{figure=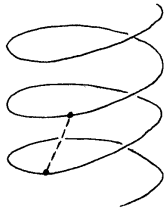,width=25mm}}}
        \caption{\small The mistake in Holst and Peld\'an: A spacelike curve may 
fail to be achronal.}
        \label{fig:2}
\end{figure}

Figueroa-O'Farrill et al. \cite{figofarr} correctly note a problem with this. Inside 
the light cones in Fig. \ref{fig:ett} a flow line of $\tilde{\xi}$ is a helix; hence 
different points on it can be connected with timelike curves. Technically 
$\tilde{\xi}$ is spacelike but not achronal, and when identifications are made along 
it closed timelike curves (CTCs) 
will arise. See Fig. \ref{fig:2}. Hence the motivation for the restricted covering 
space chosen by Holst and Peld\'an has disappeared.   

According to us the covering space of the physical spacetime should 
not be restricted in any way, and the covering space of \scri \ is restricted only 
if the isometries have fixed points there. For spacetimes of type Id there are 
no fixed points to limit the future of \scri, and therefore these spacetimes are 
not black holes. But they are rather striking counterexamples to at least some 
versions of the chronology 
protection conjecture \cite{Hawking}. The Poincar\'e ball at $t = 0$ is 
transformed into itself by $\tilde{\xi}$, hence no CTC can pass it after taking the 
quotient. Moreover the data imposed on \scri , to the future of this spatial 
slice, seem quite reasonable. Nevertheless CTCs are created by the time evolution. 

Figueroa-O'Farrill et al. draw a very different conclusion. They 
restrict the covering space even further, to the region where $J_{ZU}$ is 
spacelike---also when $a \neq 0$. The resulting quotient space is free of 
CTCs, and it is a black hole spacetime. 
From our point of view this restriction by hand is inadmissible 
because the resulting $\scri $ admits a regular completion at future infinity. 
To support our view, let us consider an experimentalist (in anti-de Sitter 
space) who plans to build a time machine, and applies for support from the 
Anti-de Sitter Research Council. Suppose that the evaluation report states 
the scheme to be impossible, because any part of spacetime containing CTCs 
will be declared void by decree of the Establishment. The experimentalist 
will rightly object to this report. Clearly she would object to the 
suggestion by Figueroa-O'Farrill et al. on the same grounds---even if their 
attitude is softened to the extent that only regions from where the Research 
Council can see the CTCs is taken away.    
It would be a different matter if the Council were able to present detailed 
arguments to show that the time machince cannot be created because of, say, 
generic small perturbations or quantum mechanical effects. Perhaps it should 
give additional funding to someone trying to work such an argument out?

Thus the conclusion, but not the argument, of Holst and Peld\'an still stands.
 
\vspace{1cm}

{\bf 5. Summary}

\vspace{5mm} 

\noindent The aim of this paper was to give an unambiguous definition of 
``BTZ black holes''. We believe we have succeeded in this. First we define 
a spacetime as the quotient of anti-de Sitter space (in any dimension) with 
a discrete isometry group. No restrictions on the covering space are needed. 
We treat \scri \ somewhat differently; if Misner singularities are present 
there, we allow the quotient to split up into components. A ``physical'' 
component is the quotient of a region where the flow of the Killing vector 
is spacelike, and it must end with a Misner singularity to the future 
(and past). We argue 
that this is a sensible procedure, in particular it is 
the only way in which to ensure that the 2+1 BTZ black hole has a component 
of its conformal boundary which is conformally equivalent to the anti-de 
Sitter \scri . 

Once this definition has been agreed on, it is seen that all the anti-de 
Sitter quotient black holes proposed in the literature do deserve their 
name---with the exception of the case under discussion, Holst's and 
Peld\'an's type Id, which is seen not to be a black hole at all. 
On the contrary it is an interesting example of a spacetime containing both 
closed timelike curves and an achronal spacelike hypersurface. 

\vspace{1cm}

{\bf Acknowledgements}

\vspace{5mm}

\noindent We thank Johan Br\"annlund for drawing our attention to the 
work of Figueroa-O'Farrill et al., and Helmut Friedrich and Paul Tod 
for some help with infinity. We also thank S\"oren Holst and 
Peter Peld\'an for comments on the manuscript---and for authorizing 
phrases like that in the caption of Fig. \ref{fig:2}.

\end{document}